\documentclass[journal]{IEEEtran}
\usepackage{upgreek}
\usepackage{graphicx}
\usepackage{subfigure}
\usepackage{amsmath,bm}
\usepackage{amssymb}

\usepackage{enumerate}

\usepackage{algorithm}
\usepackage{algpseudocode}

\usepackage[square, comma, sort&compress, numbers]{natbib}

\usepackage{color}

\newcommand{\mrr}{\color{black}}

\begin{document}
	
	\title{\huge Optimization of A Mobile Optical SWIPT System With Asymmetric Spatially Separated Laser Resonator}
	\author{
		Mingliang Xiong, Qingwen Liu,~\IEEEmembership{Senior Member,~IEEE}, and Shengli Zhou~\IEEEmembership{Fellow,~IEEE}
		
		\thanks{
			The corresponding author: Qingwen Liu.
		}
	\thanks{This work was supported in part by the National Natural Science Foundation of China under Grant 62071334, in part by the National Key Research and Development	Project under Grant 2020YFB2103900 and Grant 2020YFB2103902, in part by the Shanghai Municipal Science and Technology Major Project under Grant 2021SHZDZX0100, and in part by the Shanghai Municipal Commission of Science and Technology Project under Grant 19511132101.}
		\thanks{
			M. Xiong and Q. Liu
			are with the College of Electronics and Information Engineering, Tongji University, Shanghai 201804, China (e-mail: xiongml@tongji.edu.cn;    qliu@tongji.edu.cn). S. Zhou is with Department of Electrical and Computer Engineering, University of Connecticut, Storrs, CT 06250, USA (e-mail: shengli.zhou@uconn.edu).}

	}
	
	\maketitle
	
	\begin{abstract}
		High-power and high-rate simultaneous wireless information and power transfer~(SWIPT) becomes more and more important with the development of Internet of Things technologies. Optical SWIPT, also known as simultaneous light information and power transfer~(SLIPT), has unique advantages such as abundant spectrum resources and low propagation divergence, compared with radio-frequency (RF) SWIPT. However, optical SWIPT faces many challenges in beam steering and receiver positioning/tracking. Resonant beams generated by spatially separated laser resonators~(SSLR) have many advantages, including high power, self-aligned mobility, and intrinsic safety. It has been proposed as the carrier of wireless charging and communication. Using resonant beams, mobile electronic devices can be remotely charged and supported with high-rate data transfer. In this paper, we propose a mobile optical SWIPT system based on asymmetric SSLR and present the system optimization procedure. We also determine the boundary of the charging power and communication rate, and discuss the trade-off between power transfer and information transfer. Numerical results show that both the charging power and the communication rate of the optimized asymmetric system are much higher than those of the symmetric system in the previous work.
	\end{abstract}
	
	\begin{IEEEkeywords}
		Resonant beam communications, resonant beam charging, distributed laser charging, laser communications, wireless power transfer, 6G mobile network.
	\end{IEEEkeywords}
	
	\section{Introduction}\label{sec:intro}
	
	\IEEEPARstart{W}{ith} the development of the Internet of Things,  high-power and high-rate simultaneous wireless information and power transfer~(SWIPT) becomes more and more important. For example, augmented reality and virtual reality headsets need more energy than common mobile devices for high performance computing, and at the same time need high communication rate to exchange the three-dimensional~(3D) holographic information~\cite{a210529.02}. Another example is the unmanned aerial vehicle~(UAV) which becomes popular in digital city and some other scenarios~\cite{a210901.03}. UAVs require high-speed channel to transfer high-definition~(HD) videos, and simultaneously need enough power to expand the flying time as long as possible. Especially in the six generation~(6G) mobile networks, charging a mobile device via radio wave or laser beam is an anticipated innovation to enable many new applications~\cite{a180820.09}.
	
	Optical SWIPT exhibits distinctive advantages which can complement the weakness of radio frequency~(RF) technologies, especially in indoor environments~\cite{a210901.04}. Light wave has extremely high frequency, which provides a broad communication band.  Besides, since light has much shorter wavelength than RF, its power can be compressed in a narrow beam (e.g., laser usually has millimeter-level beam radius) to reduce the path loss. The beam width in RF beamforming is  proportional to the wavelength and  inversely proportional to the number of antenna elements and the element separation~\cite{ASPLUND202089}. As the element separation should be greater than half of the wavelength, it brings a limitation to the element number with a given antenna area and thus prompts the requirement of shortening the wavelength. Motivated by this, simultaneous light information and power transfer~(SLIPT) is receiving more and more attention~\cite{a210901.05}. 
	
	\begin{figure}[t]
		\centering
		\includegraphics[width=2.8in]{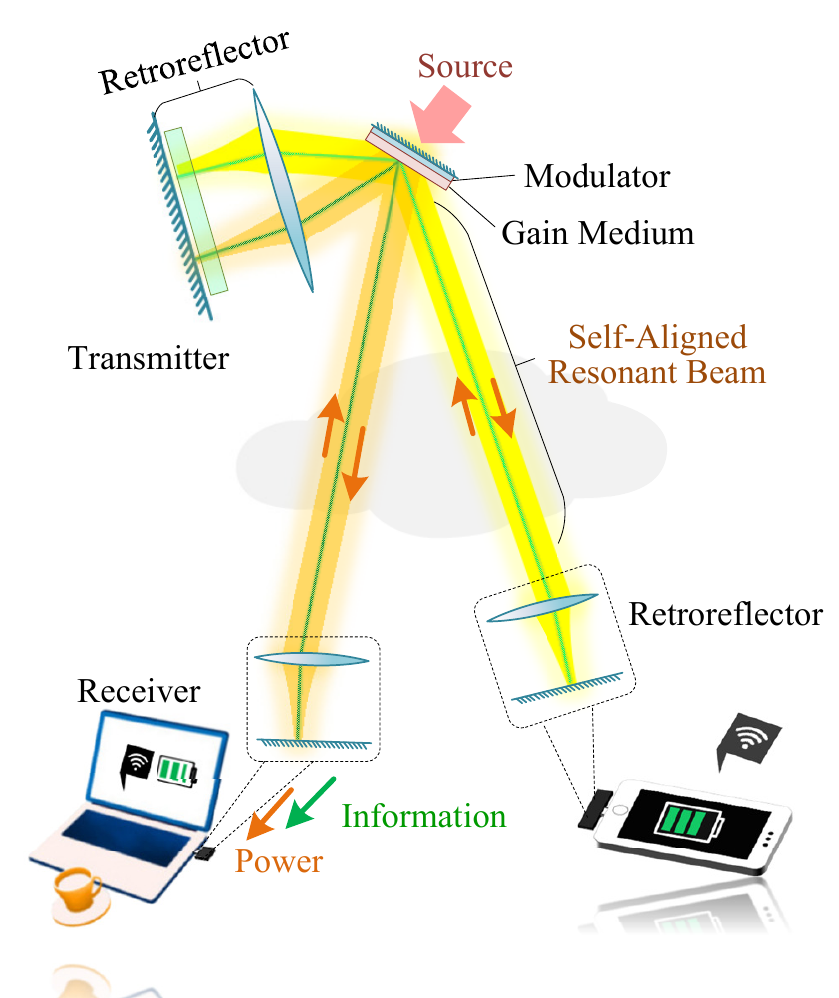}
		\caption{Scenario and mechanism of mobile optical SWIPT using self-aligned resonant beam}
		\label{fig:appli}
	\end{figure}
	
	Nevertheless, optical SWIPT still faces many challenges. For instance, the received power and the coverage area are two conflicting aspects. Visible light communications~(VLC) technologies aim to cover the whole room by emitting the light wave to everywhere~\cite{a200228.03}. In this case, the optical power received by the receiver is very low, as most power is wasted in the space. On the other hand, although the light can be focused on the receiver to improve the received power as high as possible, it faces challenges in beam steering and receiver positioning/tracking, especially for long-range applications~\cite{a220209.02}. Many works paid attention to the response speed of beam steering. For instance, in-fiber diffraction gratings employed for beam steering exhibit faster response than micro-electromechanical systems (MEMS)~\cite{a201130.03}. Two-dimensional~(2D) fiber arrays~\cite{a191111.01}, crossed gratings~\cite{a190611.03}, and silicon optical phased arrays (OPA) are also good technologies for fast beam steering~\cite{a201201.02}. Besides, beam steering with spatial light modulator (SLM) is an effective way to realize point-to-multipoint mobile optical communications~(MOC)~\cite{a190514.01}.
	However, the angle resolution of beam steering limits the acceptable distance of targets. The alignment and tracking also rely on the accuracy and the processing speed of existing positioning technologies~\cite{a210901.06}.
	
	Resonant beam has been proposed as the carrier of wireless charging and communication for its advantages including high power, self-aligned mobility, and intrinsic safety~\cite{a180727.01,a191111.06}. As shown in Fig.~\ref{fig:appli}, the resonant beam is generated by a spatially separated laser resonator~(SSLR) which consists of two retroreflectors -- one at the transmitter and the other at the receiver. The retroreflectors can be  corner cubes, cat's eyes, or telecentric cat's eyes. They have the ability to reflect the incident lights back to the source. Lights can oscillate between two retroreflectors with many round trips. If a gain medium is placed inside the resonator, the oscillating lights will be amplified, and then, form an intra-cavity  resonant beam. 
	
	The primary idea of using two corner cubes to create a very long laser~(up to $30$~km) was proposed in~\cite{a190318.02}. Resonant beam communications~(RBCom) based on telecentric cat's eye retroreflector~(TCR) is proposed in~\cite{MXiong2021}. Paper~\cite{MXiong2021} also proposed a focal TCR~(FTCR) design and demonstrated that resonators based on FTCRs can reach a stable regime where the intra-cavity diffraction loss is extremely low. The light-field simulation of the SSLR was conducted to verify its mobility~\cite{MLiu2021}. The safety of the resonant beam has also been analyzed through light-field simulation~\cite{WFang2021}. An experiment on charging a smart phone via the resonant beam generated from the FTCR-based SSLR was demonstrated in~\cite{liu2021mobile}, which achieved above $5$-W received optical power and $0.6$-W battery charging power within $2$-m distance, and also with a maximum horizontal moving range of $\pm 18$~cm. Besides, Wang \emph{et~al.} also conducted experiments on the TCR-based SSLR and demonstrated the adjustable-free
	range of $\pm 13^\circ$~\cite{a210831.01}. Lim~\emph{et~al.} proposed and did  experiments on a new SSLR cavity based on spatial wavelength division and diffraction gratings, which realized $1.7$-mW received power at $1$-m distance~\cite{a190926.02}. Liu~\emph{et~al.} expanded the field of view~(FOV) of the SSLR receiver to $\pm 30^\circ$ over $5$-m working distance in experiment~\cite{a220219.01}. Capacity of the RBCom system based on corner cubes was analyzed in~\cite{a210901.01}. To overcome the intra-cavity echo interference, an intra-cavity second harmonic generation~(SHG) scheme was adopted in the RBCom system~\cite{MXiong2021.InSHG}. 
	
	Mobile optical SWIPT based on symmetric SSLR and intra-cavity SHG was proposed in~\cite{MXiong2021.SWIPT}. However, the symmetric SSLR exhibits relatively low energy efficiency, as the beam waist locates at the midpoint of the resonator rather than the location of the gain medium. Only with an asymmetric SSLR, can the beam waist be moved to the position of the gain medium.  But, an asymmetric SSLR is really different from what we have studied in the previous work, as many parameters need to be determined. How can we create an asymmetric SSLR with proper parameters for better SWIPT performance than the previous design in~\cite{MXiong2021.SWIPT}? -- This question motivated this research.
	
	The contributions of this work are as follows.
	\begin{enumerate}
		\item[\bf 1)] We propose the  asymmetric SSLR-based mobile optical SWIPT system to provide a performance improvement compared with the symmetric system in the previous work. In the asymmetric SSLR, the two retro-reflectors have different parameters, so that the intra-cavity beam waist can be much closer to the gain medium than the symmetric SSLR whose beam waist is assured locating at the midpoint of the resonator, which improves the energy conversion efficiency of the gain medium.
		\item[\bf 2)] Since the asymmetric SSLR has several undetermined parameters, we  present a procedure to obtain the optimum parameters for the system. Using the optimization procedure, we determine the region and the boundary of the achievable charging power and achievable rate. We also discuss the trade-off between power transfer and information transfer.
	\end{enumerate}
	
	The remainder of this paper is organized as follows. In Section~\ref{sec:system}, we propose the system model of the asymmetric SSLR-based mobile optical SWIPT system.
	All the theory are presented in this Section~\ref{sec:system}. Then, we use two separated sections, Section~\ref{sec:optim} and Section~\ref{sec:result}, to present the optimization and the trade-off procedures, respectively, as they have different purposes and should be conducted in order. Specifically, in Section~\ref{sec:result} we demonstrate the performance boundary and discuss the trade-off on charging power and communication rate. The performance improvement is also verified in Section~\ref{sec:result}. At last, we conclude in Section~\ref{sec:con}.

	\begin{figure*}[t]
		\centering
		\includegraphics[width=5.6in]{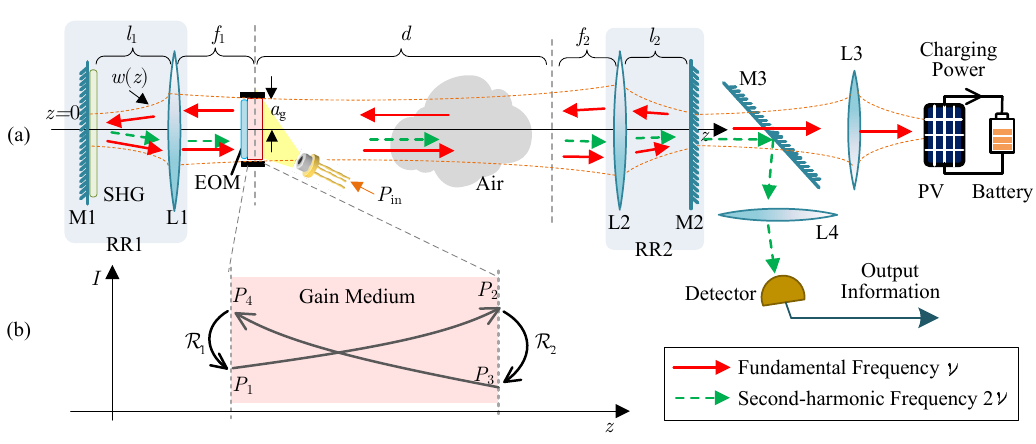}
		\caption{System description: (a) the structure of the spatially separated laser resonator and the branches of power receiving and information detection; (b)  the equivalent simplest resonator of the proposed system and the process of the intra-cavity power circulation. (Mirror M1 is high-reflective; mirror M2 is partially reflective at fundamental frequency $\nu$ with reflectivity of $R_{\rm M2}^{(\nu)}$, while it is anti-reflective at the second-harmonic frequency $2\nu$; dichroic mirror M3 split lights according to the frequency; L1 to L4 are lenses; RR1 is the retroreflector at the transmitter, consisting of M1 and L1; RR2 is the retroreflector at the receiver, consisting of M2 and L2; SHG: second harmonic generator; EOM: electro-optic modulator; PV: photovoltaic panel)}
		\label{fig:design}
	\end{figure*}

	\section{System Model}
	\label{sec:system}
	
	
	The mobile optical SWIPT system is based on the asymmetric SSLR structure and the intra-cavity SHG scheme. The resonant beam generated by the SSLR is employed as the power transfer carrier; and simultaneously, a small portion of the resonant beam is converted into a frequency-doubled SHG beam for information transfer.
	For better understanding of this work, we first describe the basic structure of the SSLR and the intra-cavity SHG scheme, although the detailed information can be found in the previous work~\cite{MXiong2021.SWIPT}.  Then, we present the system model of the asymmetric structure and point out the parameters that can be optimized.
	
	The mobile ability of this SWIPT system is supported by the SSLR, specifically by the retroreflectors in the SSLR. As is known, a typical laser cavity is comprised of two spherical mirrors that are set parallel to each other. According to the theory of resonator presented in \cite{a181221.01}, photons captured by a stable cavity will oscillate many times between two mirrors. The loss of photons is compensated for by the amplification of the gain medium  between the mirrors. Similarly, the SSLR can capture photons and force them to oscillate between two retroreflectors~(RR1 and RR2), as demonstrated in Fig.~\ref{fig:design}(a).
	Retroreflectors can reflect rays to their sources, only if the source locates in the FOV of the retroreflector. If a resonator consists of two retroreflectors,
	rays from one retroreflector can always be reflected back by the other retroreflector, and vise verse (verified by Fig. 3 in \cite{liu2021mobile}). Therefore, different from the mirrors in the typical laser cavity which need to  be in parallel to each other, the direction and the location of the retroreflectors in the SSLR are flexible. Papers \cite{MLiu2021,liu2021mobile,a210831.01} have theoretically and experimentally demonstrated the mobility/self-alignment of such double-retroreflector SSLR. Owing to this feature, the receiver can move flexibly in the FOV of the transmitter.
	
	\subsection{Ray-transfer Matrix and Beam Radius}
	In this work, we use TCRs as the components of the SSLR. A TCR consists of a plane mirror and a convex lens whose focal length is $f$; they are set in parallel to each other with a space interval of $l$. By choosing different $l$, the TCR exhibits different characteristic. Generally, we can use matrix optics theory to analysis an optical system. Each optical element can be described by a ray-transfer matrix~(see Table I in~\cite{MXiong2021.InSHG}). An optical system which consists of many elements can be expressed by multiplying the matrices of these elements (cf. Chapter~1 in~\cite{a181221.01}). For a typical TCR that has $l=f$, its ray-transfer matrix  is expressed as~\cite{a181214.03}
	\begin{align}
		\mathbf{M}_{\rm TCR}&=			\begin{bmatrix}
			1&f\\0& 1
		\end{bmatrix}
		\begin{bmatrix}
			1&0\\-1/f& 1
		\end{bmatrix}
		\begin{bmatrix}
			1&f\\0& 1
		\end{bmatrix}
		\begin{bmatrix}
			1&0\\0& 1
		\end{bmatrix}
		\begin{bmatrix}
			1&f\\0& 1
		\end{bmatrix}
		\nonumber
		\\
		&~~~~
		\begin{bmatrix}
			1&0\\-1/f& 1
		\end{bmatrix}
		\begin{bmatrix}
			1&f\\0& 1
		\end{bmatrix}
		\nonumber
		\\
		&=
		\begin{bmatrix}
			-1&0\\0 &-1
		\end{bmatrix}.
	\end{align}
	Hence, for this kind of TCR, the output ray $[r_{\rm o}, \alpha_{\rm o}]^{T}$ is  parallel to the input ray $[r_{\rm i}, \alpha_{\rm i}]^{T}$; namely
	\begin{equation}
		\begin{bmatrix}
			r_{\rm o}\\
			\alpha_{\rm o}
		\end{bmatrix}
		=\mathbf{M}_{\rm TCR}
		\begin{bmatrix}
			r_{\rm i}\\
			\alpha_{\rm i}
		\end{bmatrix} =\begin{bmatrix}
			-r_{\rm i}\\-\alpha_{\rm i}
		\end{bmatrix},
		\label{equ:rayout}
	\end{equation}
	where $r_{\rm i}$ ($r_{\rm o}$) is the displacement of the input  (output) ray relative to the optical axis at the input (output) plane; and $r_{\rm i}$ ($r_{\rm o}$) is the slope of the input (output) ray. 
	
	However, SSLRs based on typical corner cube retroreflectors or TCRs perform like plane-parallel resonators and thus exhibit very high intra-cavity diffraction loss~\cite{a210823.01,MLiu2021}. Paper~\cite{MXiong2021} found the focusing ability of the FTCR (i.e., the TCR with $l>f$) and originally proposed that FTCRs can be employed to create stable resonators which have extremely low diffraction loss. This characteristic can be recognized from the following rearranged FTCR ray-transfer matrix, as it can be viewed as the combination of a focal lens $\mathbf{M}_{\rm eqL}$ and a conventional retroreflector $\mathbf{M}_{\rm TCR}$; that is~\cite{MXiong2021}
	\begin{align}
		\mathbf{M}_{\rm FTCR}&=
		\begin{bmatrix}
			1&f\\0& 1
		\end{bmatrix}
		\begin{bmatrix}
			1&0\\-1/f& 1
		\end{bmatrix}
		\begin{bmatrix}
			1&l\\0& 1
		\end{bmatrix}
		\begin{bmatrix}
			1&0\\0& 1
		\end{bmatrix}
		\begin{bmatrix}
			1&l\\0& 1
		\end{bmatrix}
		\nonumber
		\\
		&~~~~
		\begin{bmatrix}
			1&0\\-1/f& 1
		\end{bmatrix}
		\begin{bmatrix}
			1&f\\0& 1
		\end{bmatrix}
		\nonumber
		\\
		&=\mathbf{M}_{\rm eqL} \mathbf{M}_{\rm TCR}
		\label{equ:FTCR}
	\end{align}
	where
	\begin{align}
		\mathbf{M}_{\rm eqL}=\begin{bmatrix}
			1&0\\-1/f_{\rm RR}& 1
		\end{bmatrix}, \mbox{and}~
		f_{\rm RR}=\dfrac{f^2}{2(l-f)}.
		\label{equ:fRR}
	\end{align}
	Here $\mathbf{M}_{\rm eqL}$ is identical to a lens matrix whose focal length is $f_{\rm RR}$. Referring to the theory of matrix optics (cf. Chapter 4~in \cite{a181228.01}), if $f_{\rm RR}>0$,  $\mathbf{M}_{\rm eqL}$ represents a convex lens. Hence, we set $l>f$, i.e., $f_{\rm RR}>0$, to enable the focusing ability of the equivalent lens. If $f_{\rm RR}<0$, $\mathbf{M}_{\rm eqL}$ represents a concave lens.
	
	The resonant beam is the intra-cavity standing wave formed by the photons (at the fundamental frequency $\nu$) oscillating between the transmitter and the receiver, as depicted in Fig.~\ref{fig:design}(a). In the SSLR cavity, the gain medium absorbs the pump light which is generated by a laser diode module driven with input electrical power $P_{\rm in}$ to obtain the optical amplification ability. The photons are consumed during oscillation and compensated by the stimulated emission occurring in the gain medium, which reaches a balance as time goes on. At the receiver, a portion of the oscillating photons are allowed to pass through the output mirror M2, and the other part is reflected back to maintain the resonance. A photovoltaic panel (PV) converts these output photons into electricity to charge the battery. To realize SWIPT, the information is modulated on the frequency-doubled SHG beam (at the second-harmonic frequency $2\nu$). The SHG beam is generated from the resonant beam by the SHG medium placed inside the resonator. This is practical as the intra-cavity SHG has been well studied~\cite{a220223.01,a220223.02,a220223.03}. The light intensity of the SHG beam is modulated by the electro-optic modulator (EOM) placed near the gain medium. Due to the frequency difference from the resonant beam, the SHG beam can be totally allowed to pass through M2 and bring information to the photon detector~(PD). As we propose a SWIPT scheme, the receiver should have the ability of splitting the  information and the power. Generally, there are many splitting methods for RF-based SWIPT, such as time switching, power splitting,   receiver separation, and antenna switching~\cite{a210620.01}. However, in this paper, based on the wide-spectrum advantage of light, we transfer power and information by different light frequencies and use dichroic mirror M3 to extract the SHG beam from the mixed beam by frequency difference. This can be realized provided that M2 and M3 are coated with  partially-reflective~(PR), anti-reflective~(AR), or high-reflective~(HR) coatings at corresponding frequencies ($\nu$ or $2\nu$). Since the beams for information transfer and power transfer are with separated paths in the receiver, we use lens L3 to focus the power transfer beam on the PV and use lens L4 to focus the information transfer beam on the PD.
	
	The asymmetric SSLR is comprised of two different FTCRs RR1 and RR2. At the transmitter, the focal length of the lens L1 is $f_1$, and the space interval between L1 and the mirror M1 is $l_1$. While at the receiver, the focal length and the lens-to-mirror interval with respect to RR2 are $f_2$ and $l_2$, respectively. We term the outer focal plane of an FTCR as its input/output~(IO) plane, and take the space interval between the IO planes of the two FTCRs in the SSLR as the transmission distance $d$. An optical resonator can be expressed by its signal-pass ray-transfer matrix~\cite{a200224.01,a190511.01}. We can use this matrix to analyze the resonator's characteristic, including the stability and the distribution of the intra-cavity beam raidus~(cf. Chapter 8 in~\cite{a181221.01}).
	Then, the single-pass ray-transfer matrix of the asymmetric SSLR yields~\cite{MXiong2021}
	\begin{align}
		\begin{bmatrix}
			A&B\\C&D
		\end{bmatrix}
		=&	\begin{bmatrix}
			1&0\\0&1
		\end{bmatrix}
		\begin{bmatrix}
			1&l_2\\0&1
		\end{bmatrix}
		\begin{bmatrix}
			1&0\\ -1/f_2&1
		\end{bmatrix}
		\begin{bmatrix}
			1&f_1+f_2+d\\0&1
		\end{bmatrix}	
		\nonumber
		\\
		&\begin{bmatrix}
			1&0\\ -1/f_1&1
		\end{bmatrix}
		\begin{bmatrix}
			1&l_1\\0&1
		\end{bmatrix}
		\begin{bmatrix}
			1&0\\0&1
		\end{bmatrix}.
		\label{equ:ABCD}
	\end{align}
	By calculating the ABCD matrix expressed by~\eqref{equ:ABCD}, we obtain the elements in the matrix as follows:
	\begin{equation}
		\left\{
		\begin{aligned}
			A&=-\dfrac{f_2}{f_1}-\dfrac{d}{f_1}+\dfrac{d l_2}{f_1 f_2},
			\\
			B&=f_1+f_2-\dfrac{l_2 (f_1 +d )}{f_2}-\dfrac{l_1 (f_2+d) }{f_1}+\dfrac{d l_1 l_2}{f_1f_2}+d,\\
			C&=\dfrac{d}{f_1f_2},\\
			D&=-\dfrac{f_1}{f_2}-\dfrac{d}{f_2}+\dfrac{dl_1}{f_1f_2}.\\
		\end{aligned}
		\right.\label{equ:A-B-C-D}
	\end{equation}
	A stable resonator means that rays oscillating in this resonator can not escape, and therefore, the diffraction loss is extremely low. The ABCD elements in~\eqref{equ:A-B-C-D} can be utilized to judge SSLR's stability and calculate the beam radius distribution along the optical axis by adopting  the method of creating an equivalent resonator which has two $g$-parameters $g_1^*$ and $g_2^*$, and a cavity length $L^*$~\cite{a200224.01,a190511.01,a181221.01}. Let $g_1^*=A$, $g_2^*=D$, we calculate the resonator's stability with the following quantity:
	\begin{equation}
		g_1^*g_2^*=\dfrac{(f_1^2 + d f_1 - d l_1) (f_2^2 + d f_2 - d l_2)}{f_1^2 f_2^2}.
	\end{equation}
	For a stable resonator, the following stable condition should be met~\cite{a181221.01}:
	\begin{equation}
		0<g_1^*g_2^*<1.
	\end{equation}
	
	Generally, there are multiple  transverse modes in a laser resonator. All these existing modes add up to form the laser beam. Among these modes, the fundamental mode TEM$_{00}$ exhibits the smallest radius. The TEM$_{00}$ mode radius at location $z$ on the optical axis is obtained by~\cite{a181221.01}
	\begin{equation}
		w_{00}(z)=\sqrt{-\dfrac{\lambda}{\pi\Im\left[1/q(z)\right]}},
	\end{equation}
	where $\lambda=c/\nu$ is the wavelength, $c$ is the light speed, and $\Im[\cdot]$ takes the imaginary part of a complex number. Here, parameter $q(z)$ records all the information of the TEM$_{00}$ mode, such as the mode radius, the radius of curvature of the constant phase surface, and the divergence angle. Let $L^*=B$, we can compute all the $q(z)$ parameters of an FTCR-based asymmetric SSLR as~\cite{MXiong2021}
	\begin{equation}
		q(z)=
		\left\{
		\begin{aligned}
			&j |L^*|\sqrt{\dfrac{g_2^*}{g_1^*(1-g_1^*g_2^*)}}+z, ~~~~~~z\in[0,z_{\rm L1}],\\
			&\frac{q(z_{\rm L1})}{{-q(z_{\rm L1})}/{f_1}+1}+(z-z_{\rm L1}),~ z\in(z_{\rm L1},z_{\rm L2}],\\
			&\frac{q(z_{\rm L2})}{{-q(z_{\rm L2})}/{f_2}+1}+(z-z_{\rm L2}),~ z\in(z_{\rm L2},z_{\rm M2}],\\
		\end{aligned}
		\right.
		\label{equ:qParam}
	\end{equation}
	where $j=\sqrt{-1}$; and $z_{\rm L1}$, $z_{\rm L2}$, and $z_{\rm M2}$ represent the location of L1, L2, and M2, respectively.
	
	The resonant beam radius is proportional to its TEM$_{00}$ mode radius. This proportion is called the beam propagation factor; and it is a constant at any location on the optical axis. Assuming the gain medium has the smallest aperture among all the devices in the resonator and most diffraction loss comes from the gain medium aperture, we can approximate the beam radius at location $z$ by~\cite{a181221.01}
	\begin{equation}
		w(z)=\dfrac{a_{\rm g}}{w_{00}(z_{\rm g})}\sqrt{-\dfrac{\lambda}{\pi\Im\left[1/q(z)\right]}},
		\label{equ:radius}
	\end{equation}
	where $a_{\rm g}$ is the radius of the gain medium aperture, and $z_{\rm g}=l_1+f_1$ is the location of the gain medium.
	
	\subsection{Power Computation}
	As shown in Fig.~\ref{fig:design}(b), the SSLR can be equivalent to the simplest resonator which is a gain medium with two mirrors attached to each side. The resonant beam consists of two parts -- the leftward-traveling wave and the rightward-traveling wave. $P_1$, $P_2$, $P_3$, and $P_4$ denote the traveling wave powers at four important stages in a circulating period. The equivalent reflectivity $\mathcal{R}_1$ depends on the SHG efficiency $\eta_{\rm SHG}$, the transmissivity of SHG medium $\Gamma_{\rm SHG}$ (without the SHG process), and the equivalent reflectivity of RR1 $\mathcal{R}_{\rm RR1}$. The equivalent reflectivity $\mathcal{R}_2$ depends on the transmissivities of the gain medium $\Gamma_{\rm g}$ and the air $\Gamma_{\rm air}$; and it also depends on the equivalent reflectivity of RR2 $\mathcal{R}_{\rm RR2}$ and the diffraction loss coefficient $\Gamma_{\rm diff}$. To calculate the output power of the fundamental frequency and the second-harmonic frequency released from M2, we first need to obtain the power, $P_4$, of the leftward-traveling part of the resonant beam. We can calculate $P_4$ by solving the following equations~\cite{MXiong2021.SWIPT}:
	\begin{equation}
		\left\{\begin{aligned}
			&P_4=\dfrac{\pi a_{\rm g}^2 I_{\rm s}}{(1+\sqrt{\dfrac{\mathcal{R}_1}{\mathcal{R}_2}})(1- \sqrt{\mathcal{R}_2 \mathcal{R}_1})}\left[\dfrac{\eta_{\rm c} P_{\rm in}}{\pi a_{\rm g}^2I_{\rm s} }  -\ln\frac{1}{\sqrt{\mathcal{R}_2\mathcal{R}_1}} \right], \\
			&\mathcal{R}_1=(1-\eta_{\rm SHG})^2\Gamma_{\rm SHG}^2 \mathcal{R}_{\rm RR1}, \\
			&\mathcal{R}_2=\Gamma_{\rm g}^2\Gamma_{\rm air}^2 \mathcal{R}_{\rm RR2}\Gamma_{\rm diff}, \\
			&\eta_{\rm SHG}=  \dfrac{8 \pi^2 d_{\rm eff}^2 l_{\rm s}^2 }{ \varepsilon_0 c \lambda^2 n_0^3}\cdot \dfrac{2P_4}{\pi w^2(0)}, 
		\end{aligned}
		\right.\label{equ:P4equs}
	\end{equation}
	where $I_{\rm s}$ is the saturation intensity related to the gain medium material, $\eta_{\rm c}$ is the combined pumping efficiency, and $P_{\rm in}$ is the input driving power. Note that if $P_4<0$, we should set $P_4=0$. The first line in~\eqref{equ:P4equs} is based on the Rigrod analysis introduced in Chapter 12 in~\cite{a181224.01}. The SHG efficiency $\eta_{\rm SHG}$ depends on the resonant beam intensity at the SHG medium and the material parameters, including the efficient nonlinear coefficient $d_{\rm eff}$, the crystal thickness $l_{\rm s}$, and the refractive index $n_0$; this theory can be learned from~Chapter 10 in~\cite{a181218.01}. Other factors involved in the SHG process include the resonant beam wavelength $\lambda$, the vacuum permeability $\varepsilon_0$, and the speed of light $c$. Generally, the diffraction loss coefficient $\Gamma_{\rm diff}$ is computed using a numerical simulation program, for example, with the Fox-Li method~\cite{MLiu2021}. For fast calculation, the approximation of the diffraction loss in a special case where all the devices are coaxially placed can be found in~\cite{MXiong2021}. The gain medium thickness is not a factor in the calculation of the intra-cavity beam power, but it affects the absorption efficiency to the pump source which is a factor in the combined pumping efficiency $\eta_{\rm c}$.
	The rightward-traveling power outputting from the gain medium can be computed by~\cite{MXiong2021.SWIPT}
	\begin{equation}
		P_2=\sqrt{\frac{\mathcal{R}_1}{\mathcal{R}_2}} P_4.
	\end{equation} 
	
	\begin{table} [t] 
		\caption{System  Parameters}
		\renewcommand{\arraystretch}{1.2}
		\centering
		\begin{tabular}{ l l l}
			\hline
			\textbf{Parameter} & \textbf{Symbol} &  \textbf{Value} \\
			\hline
			Saturation intensity & $I_{\rm s}$ &  $1.1976\times10^7$~W/m$^2$\\
			Resonant beam wavelength & $\lambda$ & $1064$ nm\\
			Combined pumping efficiency& $\eta_{\rm c}$ & $43.9\%$\\
			
			Efficient nonlinear coefficient & $d_{\rm eff}$ & $4.7$~pm/V\\
			Refractive index of SHG medium&$n_0$ & 2.23 \\
			PV's responsivity&$\rho$ &$0.6$~A/W \\
			Reverse saturation current& $I_0$ & $0.32~\mu$A\\
			Shunt resistance& $R_{\rm sh}$ &$53.82$~$\Omega$ \\
			Series resistance & $R_{\rm s}$ & $37$~m$\Omega$\\
			Diode ideality factor& $n$ & $1.48$\\
			Number of cells in PV panel& $n_{\rm s}$ & $1$\\
			Temperature& $T$ & $298$~K \\
			PD's responsivity& $\gamma$ & $0.4$~A/W\\
			Load resistor at the PD& $R_{\rm IL}$& $10$~k$\Omega$\\
			\hline
		\end{tabular}
		\label{tab:param}
	\end{table}
	
	The optical devices in the resonator  absorb, reflect, or refract a small proportion of the passing beam, resulting in  transmission loss. The air is also a loss factor in transmission, since the vapors and particles  scatter/absorb the lights. As presented in~\cite{MXiong2021.SWIPT}, the received optical power from both the power transfer~(PT) branch and the information transfer~(IT) branch are computed as follows. 
	The received optical power at the PV is derived from the rightward-traveling power, $P_2$, outputting from the gain medium; that is
	\begin{equation}
		P_{\rm recv,PT} = \Gamma_{\rm t}^{(\nu)} P_2,
	\end{equation}
	where $\Gamma_{\rm t}^{(\nu)}$ is the transmission attenuation coefficient for the PT beam (at the fundamental frequency $\nu$), namely
	\begin{equation}
		\Gamma_{\rm t}^{(\nu)}=\Gamma_{\rm PV}\Gamma_{\rm L3}  \Gamma_{\rm M3}^{(\nu)} \Gamma_{\rm M2}^{(\nu)} \Gamma_{\rm L2} \Gamma_{\rm air},
	\end{equation}
	where $\Gamma_{\rm PV}$ and $\Gamma_{\rm L3}$ are the transmissivities of the PV's incident surface and L3, respectively; and $\Gamma_{\rm M3}^{(\nu)}$ and $\Gamma_{\rm M2}^{(\nu)}$ are the transmissivities of M3 and M2 at frequency $\nu$, respectively.
	
	Similarly, as we have obtained the leftward-traveling power $P_4$, the received optical power at the PD yields
	\begin{equation}
		P_{\rm recv,IT} =2\eta_{\rm SHG}\Gamma_{\rm t}^{(2\nu)} P_4.
	\end{equation}
	Here, the transmission attenuation coefficient, $\Gamma_{\rm t}^{(2\nu)}$, for the IT beam~(at the second-harmonic frequency $2\nu$) is 
	\begin{equation}
		\Gamma_{\rm t}^{(2\nu )}=\Gamma_{\rm det}\Gamma_{\rm L4}R_{\rm M3}^{(2\nu)} \Gamma_{\rm M2}^{(2\nu)}\Gamma_{\rm L2}\Gamma_{\rm air}\Gamma_{\rm g,EOM}\Gamma_{\rm L1},
	\end{equation}
	where $\Gamma_{\rm det}$, $\Gamma_{\rm L4}$, $\Gamma_{\rm L2}$, $\Gamma_{\rm L1}$, and $\Gamma_{\rm g,EOM}$ are the transmissivities of the PD's incident surface, L4, L2, L1, and the combined body of the gain medium and the EOM, respectively; $R_{\rm M3}^{(2\nu)}$ is the reflectivity of the dichroic mirror M3  at frequency $2\nu$; and  $\Gamma_{\rm M2}^{(2\nu)}$ is the transmissivity of mirror M2 at frequency $2\nu$. 
	
	The PT beam with optical power $P_{\rm recv,PT}$ is received by the PV and then converted into electricity for battery charging. The charging current $I_{\rm chg}$ is expressed as~\cite{a190923.01,a180802.01}
	\begin{equation}
		\left\{
		\begin{array}{lr}
			I_{\rm chg}=\rho P_{\rm recv,PT}-I_{\rm d}-\dfrac{V_{\rm d}}{R_{\rm sh}}, \vspace{1ex}\\
			I_{\rm d}=I_0 \left[ \exp\left({\dfrac{V_{\rm d}}{n_{\rm s}n V_{T}}}\right) -1 \right],  \vspace{1ex}\\
			V_{\rm d}=I_{\rm chg}(R_{\rm PL}+R_{\rm s}), \vspace{1ex}\\
			V_{T}=\dfrac{kT}{\mathsf{e}},
		\end{array}
		\right. \label{equ:PV}
	\end{equation}
	where $\rho$ is the responsivity of PV, $R_{\rm sh}$ is the internal equivalent shunt resistor, $R_{\rm s}$ is the internal equivalent series resistor, $I_0$ is the reverse saturation current, $n_{\rm s}$ is the number of cells in the PV module, $n$ is the ideality factor of the internal equivalent diode, $R_{\rm PL}$ is the equivalent load resistor, 
	$k$ is Boltzmann's constant, $T$ is the temperature, and $\mathsf{e}$ is the electron charge. The charging power, $P_{\rm chg}$, which outputs from the PV  also depends on the charging voltage $V_{\rm chg}$ on the output port. In practice, a maximum power point tracking~(MPPT) circuit is connected between the PV and the battery to achieve the maximum PV conversion efficiency. In theory, employing the equivalent circuit model expressed in~\eqref{equ:PV}, we can obtain the charging power at the maximum power point by solving the following problem~\cite{MXiong2021.SWIPT}:
	\begin{align}
		\textrm{P1}:~~P_{\rm chg} =\max\limits_{V_{\rm chg}}~& I_{\rm chg} V_{\rm chg}, \vspace{2ex}\\
		\mbox{ s.t.}~~
		&R_{\rm PL}=\dfrac{V_{\rm chg}}{I_{\rm chg}},\vspace{1ex}\\
		&0\leqslant V_{\rm chg}~\leqslant V_{\rm oc},
		\label{equ:Pchg-max-st}
	\end{align}
	where  $V_{\rm oc}$ is the open-circuit voltage of the PV. 
	
	In the IT branch, the SHG beam with optical power $P_{\rm recv,IT}$ is received at the PD. For such an intensity modulation channel, the achievable rate is obtained as~\cite{a211104.01,MXiong2021.SWIPT}
	\begin{equation}
		R_{\rm b}=\dfrac{1}{2}\log_2\left\{1+\dfrac{(\gamma P_{\rm recv,IT})^2}{2\pi  e\sigma_{\rm n}^2}\right\},
	\end{equation}
	where $e$ is the nature constant, and $\sigma_{\rm n}^2$ is the variance of the noise given by~\cite{a200427.02}
	\begin{equation}
		\sigma_{\rm n}^2=2\mathsf{e}(\gamma P_{\rm recv,IT} + I_{\rm bk}) B+\frac{4kT B}{R_{\rm IL}},
	\end{equation}
	where $\gamma$ is the responsivity of the PD, $I_{\rm bk}= 5100~\mu$A is the background radiation-induced photon current, $B=800$~MHz is the bandwidth, and $R_{\rm IL}=10~\mbox{k}\Omega$ is the load resistance.
	
	According to the above analysis, we find that many parameters should be optimized, including the focal lengths of the lenses, the mirror-to-lens space intervals of the FTCRs,  the gain medium radius, and the reflectivity of the output mirror M2. Besides, the SHG medium thickness should be considered as a factor in the power allocation of PT and IT, as it determines the SHG efficiency. In the following section, we  detail the system optimization procedure.
	
	\section{System Optimization}
	\label{sec:optim}
	
	Section~\ref{sec:system} has presented the system model and the system parameters which can be adjusted to optimize the system performance. Here we pursue an optimization problem to maximize the transmission efficiency defined as 
	\begin{equation}
		\eta_{\rm trans}^\star=\mathop{\max}_{\mathbf{v},\mathbf{m}}\dfrac{P_{\rm recv,PT}+P_{\rm recv,IT}}{P_{\rm in}},
	\end{equation}
	where $\mathbf{v}:=(l_1, f_1, l_2, f_2)$ represents the cavity structure parameter tuple that affects the resonator's stability and the TEM$_{00}$ mode radius; and $\mathbf{m}:=(a_{\rm g},  R_{\rm M2}^{(\nu)})$ is the functional parameter tuple that affects the power loss in the power-circulating process. Here $R_{\rm M2}^{(\nu)}=1-\Gamma_{\rm M2}^{(\nu)}$ is the reflectivity of M2 at the fundamental frequency. The decision of the SHG medium thickness $l_{\rm s}$ should be considered, but it is not an optimization parameter (will be justified in Section~\ref{sec:result}). We choose $l_{\rm s}$ relying on the trade-off between PT and IT. Other parameters such as the gain medium material and thickness, the SHG medium material, and the wavelength are not the candidate parameters to be optimized, since the selection of them depends on not only the performance but also the requirements. Parameters such as $I_{\rm s}$ and $\eta_{\rm c}$ should be as large as possible, but these parameters depend on the selection of the material or the manufacture of the pump module.
	
	We first optimize the cavity structure parameters to obtain a smallest TEM$_{00}$ mode radius at the gain medium. This optimization supports us to set the gain medium radius as small as possible, which can improve the light amplification ability of the gain medium under the same input source power. Then, two functional parameters $a_{\rm g}$ and $R_{\rm M2}^{(\nu)}$ are optimized to reduce the power loss in the middle process of the intra-cavity power circulation. Unless otherwise specified, the  parameters to be used can be found in Table~\ref{tab:param}. Most parameters values are identical to those in~\cite{MXiong2021.SWIPT}, including the material parameters for the gain medium, the SHG medium, the PV and the PD. The lenses and mirrors still have absorption, reflection, or refraction at their surface even if they are coated with AR or HR coatings. Hence, we set $\{\Gamma_{\rm L1}, \Gamma_{\rm L2}, \Gamma_{\rm L3},  \Gamma_{\rm L4}, \Gamma_{\rm SHG}, \Gamma_{\rm M2}^{(2\nu)}, \Gamma_{\rm M3}^{(\nu)}\} = 99\%$, $\{\Gamma_{\rm det}, \Gamma_{\rm PV}\}=99.5\%$, $\{R_{\rm M1}, R_{\rm M3}^{(2\nu)}\}=99.5\%$, $\Gamma_{\rm g}=98.51\%$, and $\Gamma_{\rm g,EOM}=97.52\%$~\cite{MXiong2021.SWIPT}. The loss induced by the air is  distance-dependent, namely $\Gamma_{\rm air}=e^{-\alpha d}$, where $\alpha=0.0001$ for clear air~\cite{a200427.04}.
	
	\subsection{Cavity Parameters}
	
\begin{figure}
		\centering
		\includegraphics[width=3.3in]{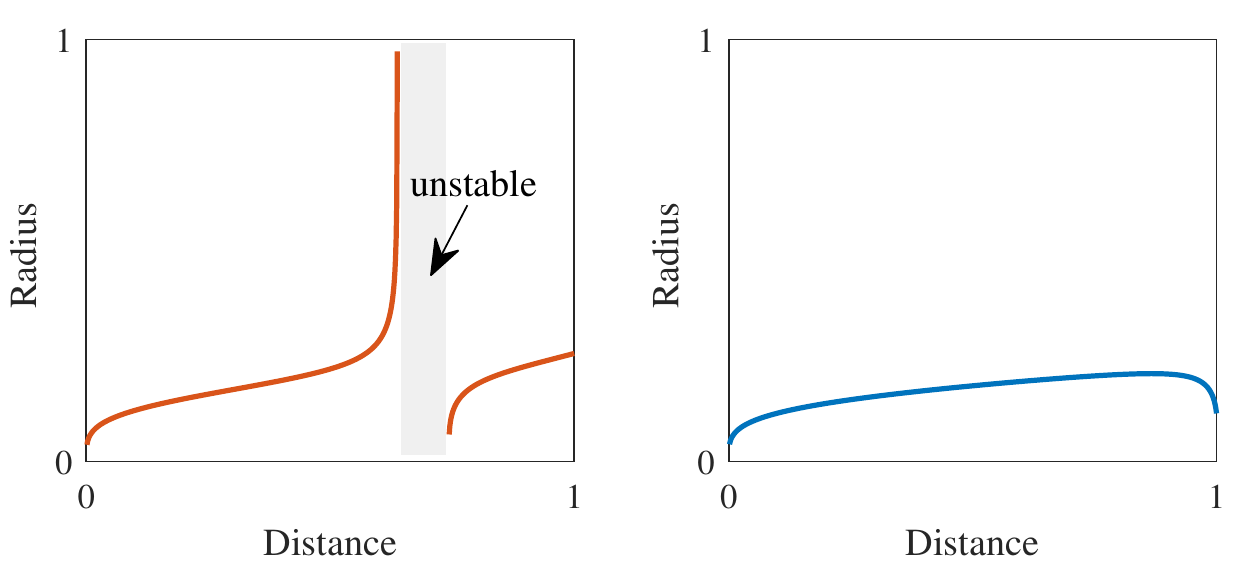}
		\caption{Diagram of the TEM$_{00}$ mode radius at the gain medium varying with the transmission distance in two cases: (a) The variation is discontinuous due the existence of unstable regime; (b) the variation is continuous}
		\label{fig:discontin}
	\end{figure}

	The cavity structure parameter tuple $\mathbf{v}$ determines the resonator stability and the TEM$_{00}$ mode radius, $w_{00}(z_{\rm g})$, at the gain medium. The intra-cavity diffraction loss depends on the ratio of the gain medium radius to the TEM$_{00}$ mode radius at the gain medium, as depicted in~\eqref{equ:radius}. Approximately, when the gain medium radius is two or three times greater than the TEM$_{00}$ mode radius, the  diffraction loss can be neglected. Consequently, given the expected distance $d_{\rm set}$, we should find the smallest $w_{00}(z_{\rm g})$ to minimize the diffraction loss. The optimum cavity structure parameters to provide the smallest $w_{00}(z_{\rm g})$ is obtained by solving the following problem:
	\begin{subequations}
		\begin{align}
			\textrm{P2}:~\mathbf{v}^{\star}=&\mathop{\arg\min}_{\mathbf{v}\in \mathbb{R}^+}~{w_{00}(z_{\rm g})}\Big|_{ d=d_{\rm set}},\\
			\mathtt{s.t.}~~
			&w_{00}(z_{\rm L1})\leqslant a_{\rm L1, bound},\\
			&w_{00}(z_{\rm L2})\leqslant a_{\rm L2, bound},\\
			&0<g_1^*g_2^*<1~ \mbox{for all}~d~ \in~[d_{\rm min},d_{\rm max}].\label{equ:v-gg}
		\end{align}
	\end{subequations}
	The optimization search should be limited in a positive real region $\mathbb{R}^+$ where all the parameters are positive real numbers. The distance is set in advance to be an expected distance $d_{\rm set}$. We also set the mode radius boundary at the lens L1 and L2 to be $a_{\rm L1,bound}$ and $a_{\rm L2,bound}$, respectively, in order to prevent any serious diffraction loss occurring at the lenses, since it will occur when the TEM$_{00}$ mode radius approaches or exceeds the devices' radius. As shown in Fig.~\ref{fig:discontin}(a), some parameters may make the cavity unstable in some distance range, which is undesired. The cavity should always be stable when the distance changes within the expected moving range~$[d_{\rm min},d_{\rm max}]$; see Fig.~\ref{fig:discontin}(b). Therefore, the condition in \eqref{equ:v-gg} should be satisfied. 
	
	We solve \textrm{P2} with the Monte Carlo method. The algorithm is depicted in Algorithm~\ref{alg:cavity-optim}. We first choose $N_{\rm s}$ samples  $\mathbf{v}_i~(i=1, 2, \dots, N_{\rm s})$ randomly as the initial parameters, and find the optimum parameter tuple $\mathbf{v}^*$; then, we choose a smaller search range around the optimum parameter $\mathbf{v}^*$, generate another series of samples  randomly within the updated search range, and find the optimum one from these new samples. By repeating the above procedures, we can obtain the optimum solution. In Algorithm~\ref{alg:cavity-optim}, $\infty$ is infinity, and $\mathsf{rand}(\mathbf{v}_{\rm LBound}, \mathbf{v}_{\rm UBound})$ is a random function which returns a random tuple $\mathbf{v}$ within the range lower-bounded by $\mathbf{v}_{\rm LBound}$ and upper-bounded by $\mathbf{v}_{\rm UBound}$. Note that each tuple has many elements, so in the pseudocode the  statements that include tuples should be expanded independently for each element in tuples. For example, Line \ref{alg:ifstart}--\ref{alg:ifend} should be expanded to four \textbf{if} statement blocks for $f_1$, $f_2$, $l_1$, and $l_2$ independently. The input parameters $N_{\rm itr}=30$, $N_{\rm smax}=1000000$. All the elements in $\mathbf{v}_{\rm LBound}$ (the lower bounds for $f_1$, $f_2$, $l_1$, and $l_2$) are set to zero. All the elements in $\mathbf{v}_{\rm UBound}$ (the upper bounds for $f_1$, $f_2$, $l_1$, and $l_2$) are set to $0.006$~m. $\alpha_{\rm sc}=0.7$ specifies a scale factor for the reduction of the search range. The remaining parameters are set as $\{d_{\rm set}, a_{\rm L1,bound}, a_{\rm L2,bound},d_{\rm min}, d_{\rm max}\}=\{6~\mbox{m},~3~\mbox{mm},~3~\mbox{mm},~0~\mbox{m},~6~\mbox{m}\}$.
	
	\begin{algorithm}[t]
		
		\caption{Optimizing Cavity Parameters}\label{alg:cavity-optim}
		\begin{algorithmic}[1]
			\Require  $N_{\rm itr}$, $N_{\rm smax}$, $\mathbf{v}_{\rm LBound,in}$, $\mathbf{v}_{\rm UBound,in}$, $d_{\rm set}$, $d_{\rm min}$, $d_{\rm max}$, $a_{\rm L1,bound}$, $a_{\rm L2,bound}$, $\alpha_{\rm sc}$
			\Ensure $\mathbf{v}^*$
			\State $w_{\rm 00,min}\gets +\infty$
			\State $\mathbf{v}_{\rm diff}\gets \mathbf{v}_{\rm UBound,in}-\mathbf{v}_{\rm LBound,in}$
			
			\State $\mathbf{v}_{\rm LBound}\gets \mathbf{v}_{\rm LBound,in}$
			
			\State $\mathbf{v}_{\rm UBound}\gets \mathbf{v}_{\rm UBound,in}$
			
			\For{$j_{\rm itr} = 1$ to $N_{\rm itr}$}
			\State $N_{\rm s}\gets N_{\rm smax}/j_{\rm itr}^2$ 
			\State initializing  $\mathbf{v}[N_{\rm s}]$ and $\mathbf{v}^*$
			\For{ $i=1$ to $N_{\rm s}$}
			\State $\mathbf{v}[i] \gets  \mathsf{rand}(\mathbf{v}_{\rm LBound}, \mathbf{v}_{\rm UBound})$
			\State $w_{\rm 00,g}\gets w_{\rm 00}(z_{\rm g})\big|_{ d=d_{\rm set}, \mathbf{v}=\mathbf{v}[i]}$
			\If{$w_{\rm 00,g}<w_{\rm 00,min}$\\
				~~~~~~~~and $w_{\rm 00}(z_{\rm L1})\big|_{ d=d_{\rm set}, \mathbf{v}=\mathbf{v}[i]}\leqslant a_{\rm L1,bound}$\\ ~~~~~~~~and $w_{\rm 00}(z_{\rm L2})\big|_{ d=d_{\rm set}, \mathbf{v}=\mathbf{v}[i]}\leqslant a_{\rm L2,bound}$\\	~~~~~~~~and $0<g_1^*g_2^*\big|_{\mathbf{v}=\mathbf{v}[i]}<1$ for $d\in[d_{\rm min}, d_{\rm max}]$}
			
			\State $w_{\rm 00,min}\gets w_{\rm 00,g}$
			\State $\mathbf{v}^*\gets \mathbf{v}[i]$
			
			\EndIf
			
			\EndFor
			
			\State $\mathbf{v}_{\rm diff}\gets \alpha_{\rm sc}\mathbf{v}_{\rm diff}$
			\State $\mathbf{v}_{\rm LBound}\gets \mathbf{v}^*-\mathbf{v}_{\rm diff}$
			\State $\mathbf{v}_{\rm UBound}\gets \mathbf{v}^*+\mathbf{v}_{\rm diff}$
			
			\If{$\mathbf{v}_{\rm LBound}<\mathbf{v}_{\rm LBound,in}$}\label{alg:ifstart}
			\State $\mathbf{v}_{\rm LBound} \gets \mathbf{v}_{\rm LBound,in}$
			\EndIf\label{alg:ifend}
			\If{$\mathbf{v}_{\rm UBound}>\mathbf{v}_{\rm UBound,in}$}
			\State $\mathbf{v}_{\rm UBound} \gets \mathbf{v}_{\rm UBound,in}$
			\EndIf
			\EndFor
		\end{algorithmic}
	\end{algorithm}
	
\begin{figure}
		\centering
		\includegraphics[width=2.5in]{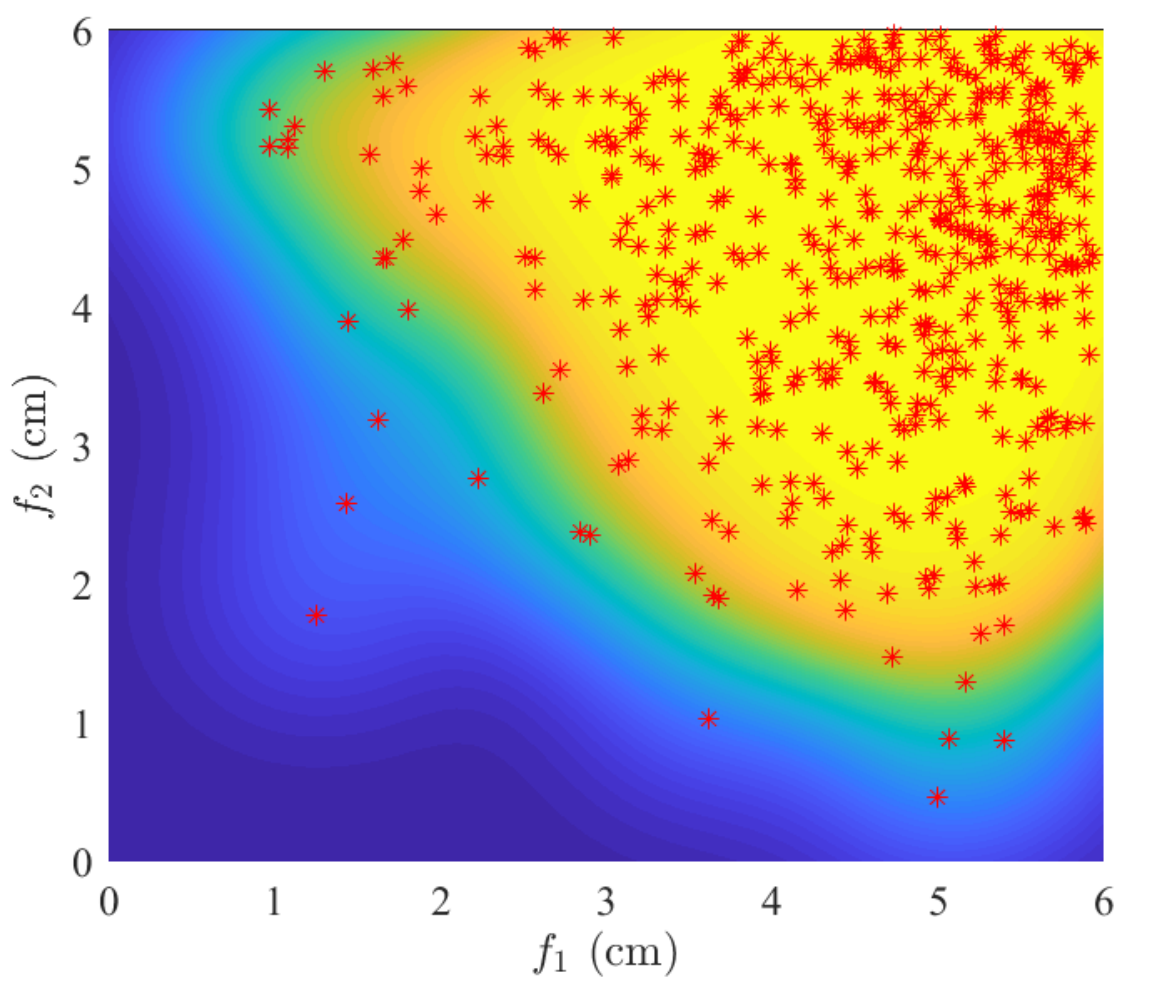}
		\caption{Distribution of the solutions of the cavity structure parameter optimization algorithm projected on the $(f_1, f_2)$ plane (Region with lighter color has higher probability of getting the solution)}
		\label{fig:f1f2}
	\end{figure}

	We executed Algorithm~\ref{alg:cavity-optim} five hundred times independently. All the solutions are projected, as points, to the $(f_1, f_2)$ plane, as shown in Fig.~\ref{fig:f1f2}. We can observe that the solutions can be obtained almost everywhere in the $(f_1, f_2)$ plane. The mode radius values $w_{00}(z_{\rm g})_m$ (for $m=1, 2, 3,\dots, 500$) computed with these  solutions are very close to each other, i.e., the average of these $w_{00}(z_{\rm g})_m$ values is $0.683$~mm while the maximum difference among these values is $0.006~$mm. This phenomenon indicates that $f_1$ and $f_2$ are not the determinants of the optimization.
	On the other hand, we can see that the solutions are not uniformly distributed in the ($f_1$, $f_2$) plane. Namely, large $f_1$ and $f_2$ are more possible to be found as the solution. This phenomenon is explained as follows. As depicted in~\eqref{equ:fRR}, the equivalent focal length, $f_{\rm RR}$, of the FTCR depends on the ratio of $f^2$ to $(l-f)$. For larger $l$ and $f$, the change of $(l-f)$ has a smaller effect on $f_{\rm RR}$; and thus, they have larger possibility to be found in the Monte Carlo method. According to the above analysis, we  conclude that the selection of $f_1$ and $f_2$ only depends on the application situation. In practice, lenses with shorter focal length have more convex surface, and therefore, exhibit higher reflectivity to the edge of the incident beam. However, a very long focal length is undesired, as the FTCR's volume will be very large.

\begin{figure}
		\centering
		\includegraphics[width=3.4in]{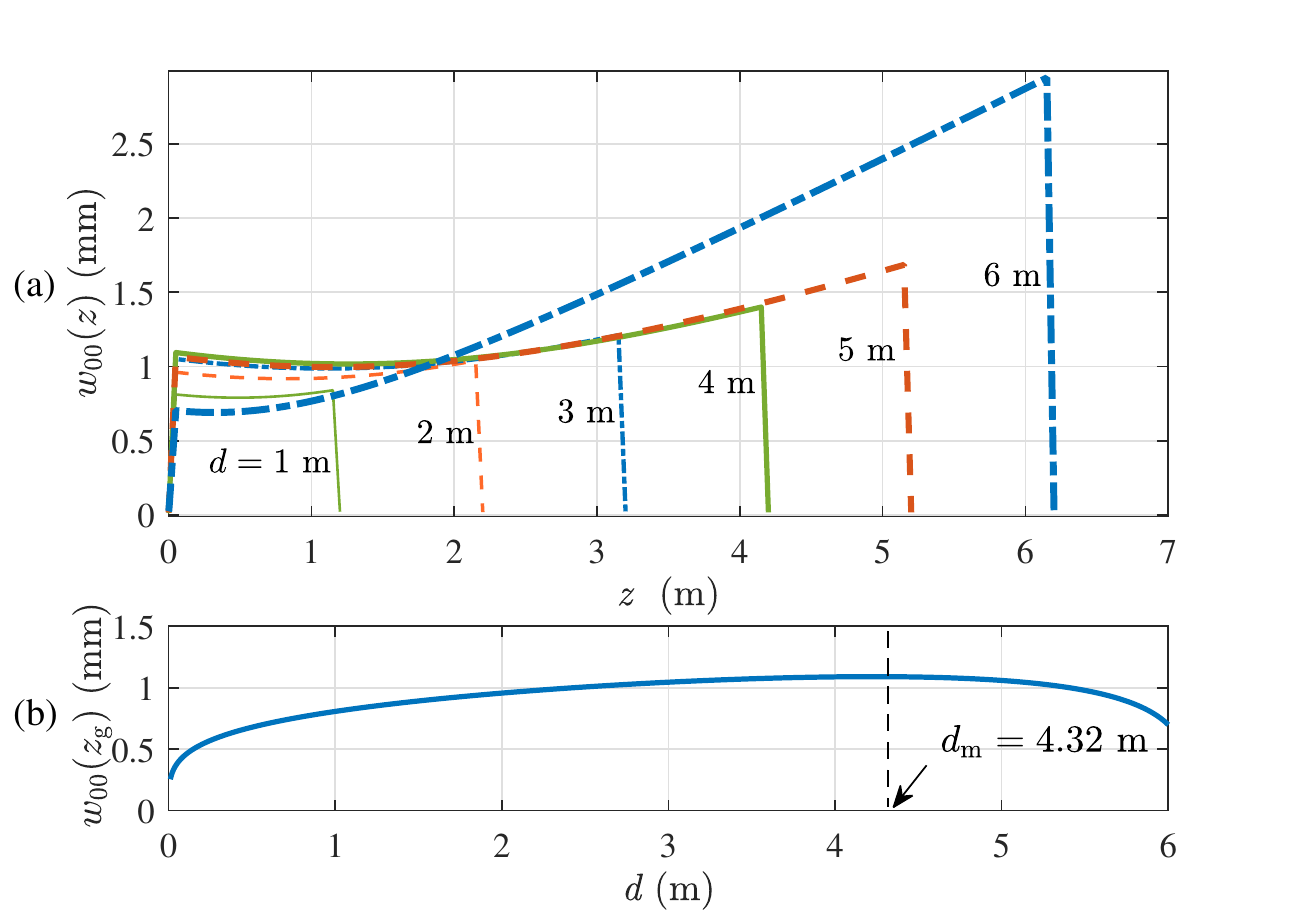}
		\caption{TEM$_{00}$ mode radius $w_{00}$ computed with the optimum cavity structure parameters obtained under $d_{\rm set}=6$~m: (a) Distribution of $w_{00}$ on the cavity axis $z$ with different transmission distance $d$; (b) $w_{00}$ at the gain medium location $z_{\rm g}$ varies with  $d$}
		\label{fig:cav-radius}
	\end{figure}

	Considering the application requirements, $f_1$ and $f_2$ should be set as the adequate values  $f_{\rm 1,set}$ and $f_{\rm 2,set}$ in advance, respectively. With fixed lens focal length, we can 
	optimize lens placement via a new optimization problem as:
	\begin{equation}
		\begin{aligned}
			\textrm{P3}:~
			(l_1^{\star},l_2^{\star})&=\mathop{\arg\min}_{(l_1,l_2)}~{w_{00}(z_{\rm g})}\Bigg|_{\substack{d=d_{\rm set}~~\\
					f_1=f_{1, \rm set}\\ f_2=f_{2,\rm set}}},\\
			\mathtt{s.t.}~~
			&w_{00}(z_{\rm L1})\leqslant a_{\rm L1, bound},\\
			&w_{00}(z_{\rm L2})\leqslant a_{\rm L2, bound},\\
			&0<g_1^*g_2^*<1~ \mbox{for all}~d~\in~[d_{\rm min},d_{\rm max}].
		\end{aligned}
	\end{equation}
	By solving  \textrm{P3} with adequate preset parameters and moving range, i.e.,   $\{d_{\rm set}, f_{\rm 1,set},f_{\rm 2,set}, a_{\rm L1,bound}, a_{\rm L2,bound}\}=\{6~\mbox{m},~5~\mbox{cm},~5~\mbox{cm},~3~\mbox{mm},~3~\mbox{mm}\}$ and $[d_{\rm min}, d_{\rm max}]=[0~\mbox{m},~6~\mbox{m}]$, we obtain the optimum values of the mirror-to-lens  intervals as
	\begin{equation}
		\left\{
		\begin{array}{l}
			l_1^\star=5.027~ \mbox{cm},\\
			l_2^\star=5.041~ \mbox{cm}.
		\end{array}
		\right.
	\end{equation}
	With the optimum cavity structure parameters, we draw the curve of  function $w_{00}(z)$  for $d=1,2,3,\dots,6$~m, as depicted in Fig.~\ref{fig:cav-radius}(a). The two turning points on the left-hand side and the right-hand side of a curve represent the TEM$_{00}$ mode radii at lenses L1 and L2, respectively. We can observe that, at the expected transmission distance, i.e., $d=d_{\rm set}$, the TEM$_{00}$ mode radius at L2, denoted by $w_{00}(z_{\rm L2})$, reaches the given boundary. As the transmission distance $d$ decreases, $w_{00}(z_{\rm L2})$ decreases correspondingly. Figure~\ref{fig:cav-radius}(b) demonstrates the variation of the TEM$_{00}$ mode radius at the gain medium $w_{00}(z_{\rm g})$ with the distance $d$. We see that $w_{00}(z_{\rm g})$ reaches the maximum value of $1.09$~mm at the distance of $4.32$~m. Here we define a distance $d_{\rm m}=4.32~$m which leads to the maximum $w_{00}(z_{\rm g})$, and thus, the highest diffraction loss within the expected moving range. The following optimization is conducted under this preset distance~$d_{\rm m}$.
	
	\subsection{Gain Medium Aperture and SHG Medium Thickness}

	The next step is to optimize the gain medium radius $a_{\rm g}$ and the SHG medium thickness $l_{\rm s}$. As mentioned before, reducing $a_{\rm g}$ leads to the increase of the diffraction loss. Nevertheless, increasing $a_{\rm g}$ results in the increase of the threshold power $P_{\rm th}$, which also reduces the transmission efficiency~$\eta_{\rm trans}$. By reformulating~\eqref{equ:P4equs}, we obtain
	\begin{equation}
		P_4=\eta_{\rm slop}\left[ P_{\rm in} - P_{\rm th} \right],
		\label{equ:P4reform}
	\end{equation}
	where
	\begin{equation}
		\eta_{\rm slop}=\dfrac{\eta_{\rm c} }{(1+\sqrt{\dfrac{\mathcal{R}_1}{\mathcal{R}_2}})(1- \sqrt{\mathcal{R}_2 \mathcal{R}_1})},
		\label{equ:eta-slop}
	\end{equation}
	and 
	\begin{equation}
		P_{\rm th}= \dfrac{\pi a_{\rm g}^2I_{\rm s}}{ \eta_{\rm c}}  \ln\dfrac{1}{\sqrt{\mathcal{R}_2\mathcal{R}_1}}.
		\label{equ:Pth}
	\end{equation}
	From~(\ref{equ:P4reform}--\ref{equ:Pth}), we observe that the threshold power $P_{\rm th}$ is proportional to $a_{\rm g}$. Consequently, to balance the intra-cavity diffraction loss and the threshold power, we should find a proper $a_{\rm g}$. Moreover, the output mirror's reflectivity $R_{\rm M2}^{(\nu)}$ also affects the power-circulating process, as it is contained in $\mathcal{R}_2$. Hence, we need to optimize $a_{\rm g}$ and $R_{\rm M2}^{(\nu)}$ concurrently to obtain a maximum transmission efficiency. As depicted in Fig.~\ref{fig:cav-radius}(b), the TEM$_{00}$ mode radius at the gain medium reaches the maximum value when the receiver is at the  distance $d_{\rm m}$. Thus, the optimization is conducted under $d=d_{\rm m}$. The reason is as follows. Optimizing $a_{\rm g}$ at this distance  brings a minimized diffraction loss to this distance; and then, the diffraction loss will be smaller for $d\neq d_{\rm m}$ due to the decreased $w_{00}(z_{\rm g})$. If the optimized $a_{\rm g}$ is obtained under other preset distance, a specific distance range around $d_{\rm m}$ will exhibit serious diffraction loss. The optimization problem is expressed as follows:
	\begin{subequations}
		\begin{align}
			\textrm{P4}:~~\mathbf{m}^\star=&\mathop{\arg\max}_{\mathbf{m}}~\eta_{\rm trans}\bigg|_{\substack{		\mathbf{v}=\mathbf{v}^\star~~\\d=d_{\rm m}~~\\
					l_{\rm s}=l_{\rm s,set}
			}}\\
			\mathtt{s.t.}~~&
			a_{\rm g}\geqslant w_{00}(z_{\rm g})\label{equ:opt-agRm2-stag}\\
			&0 \leqslant R_{\rm M2}^{(\nu)}\leqslant 1
		\end{align}
	\end{subequations}
	By solving \textrm{P4}, we can obtain the optimum functional parameter tuple $\mathbf{m}^\star:=(a_{\rm g}^\star, R_{\rm M2}^{(\nu)\star})$. Here we give an expected SHG medium thickness $l_{\rm s, set}$. This parameter affects the power allocation of the PT beam and  the IT beam, which is discussed in the next section. As depicted in~\eqref{equ:opt-agRm2-stag}, $a_{\rm g}$ is restricted to be greater than $w_{00}(z_{\rm g})$ because a smaller gain medium aperture will lead to serious  diffraction loss. Besides, the reflectivity $R_{\rm M2}^{(\nu)}$ should be in the range between 0 and 1.

\begin{figure}
		\centering
		\includegraphics[width=3.4in]{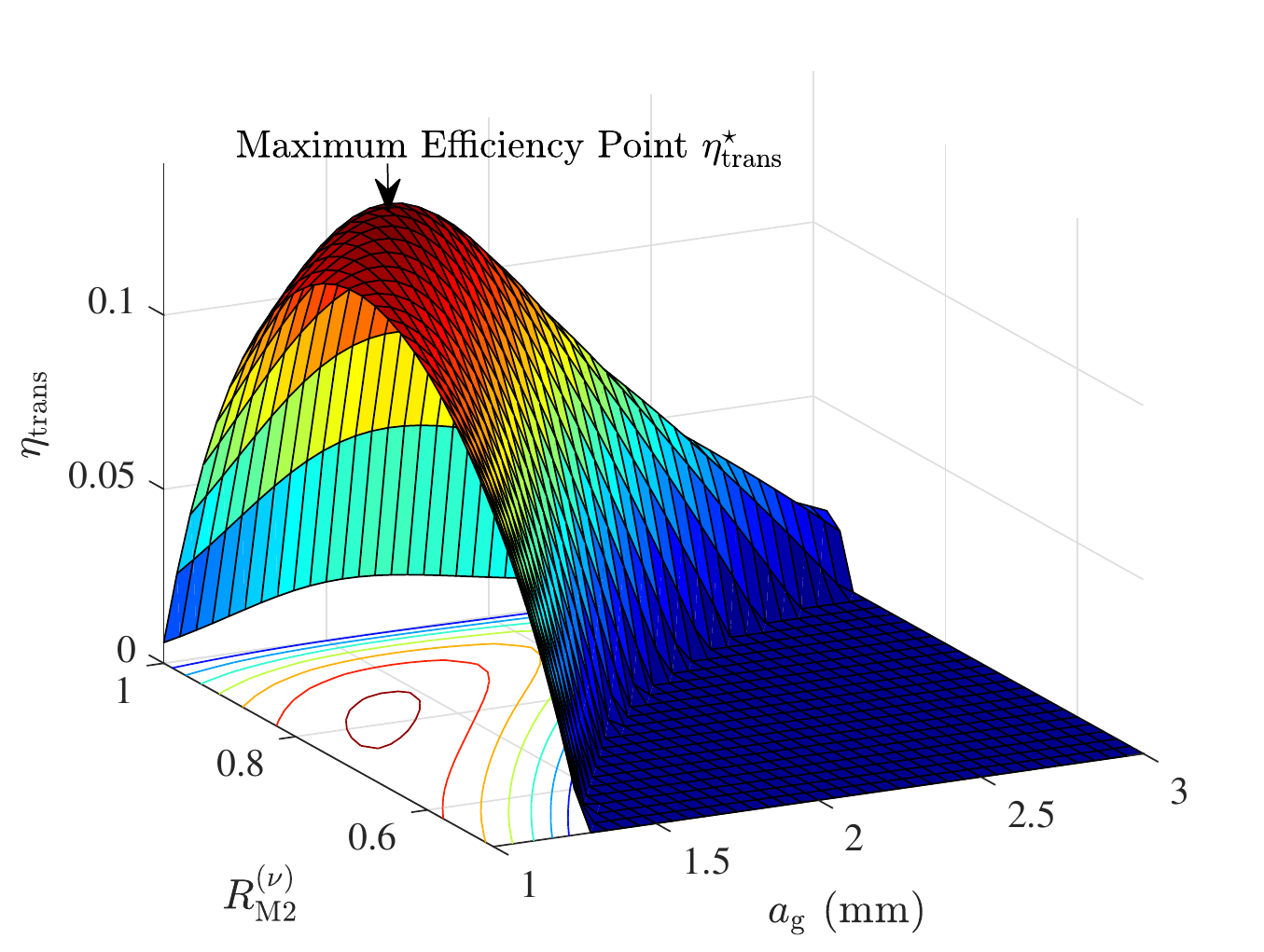}
		\caption{Transmission efficiency $\eta_{\rm trans}$ under different reflectivity, $R_{\rm M2}^{(\nu)}$, of the output mirror M2 and gain medium radius $a_{\rm g}$ (The driving power $P_{\rm in}=60$~W; the SHG medium thickness $l_{\rm s}=0.5$~mm; the transmission distance $d=4.32$~m; the focal lengths $\{f_1,f_2\}=5$~cm; the mirror-to-lens intervals $\{l_1, l_2\}$ are obtained by optimization procedure)}
		\label{fig:surfc}
	\end{figure}

\begin{figure}
		\centering
		\includegraphics[width=3.4in]{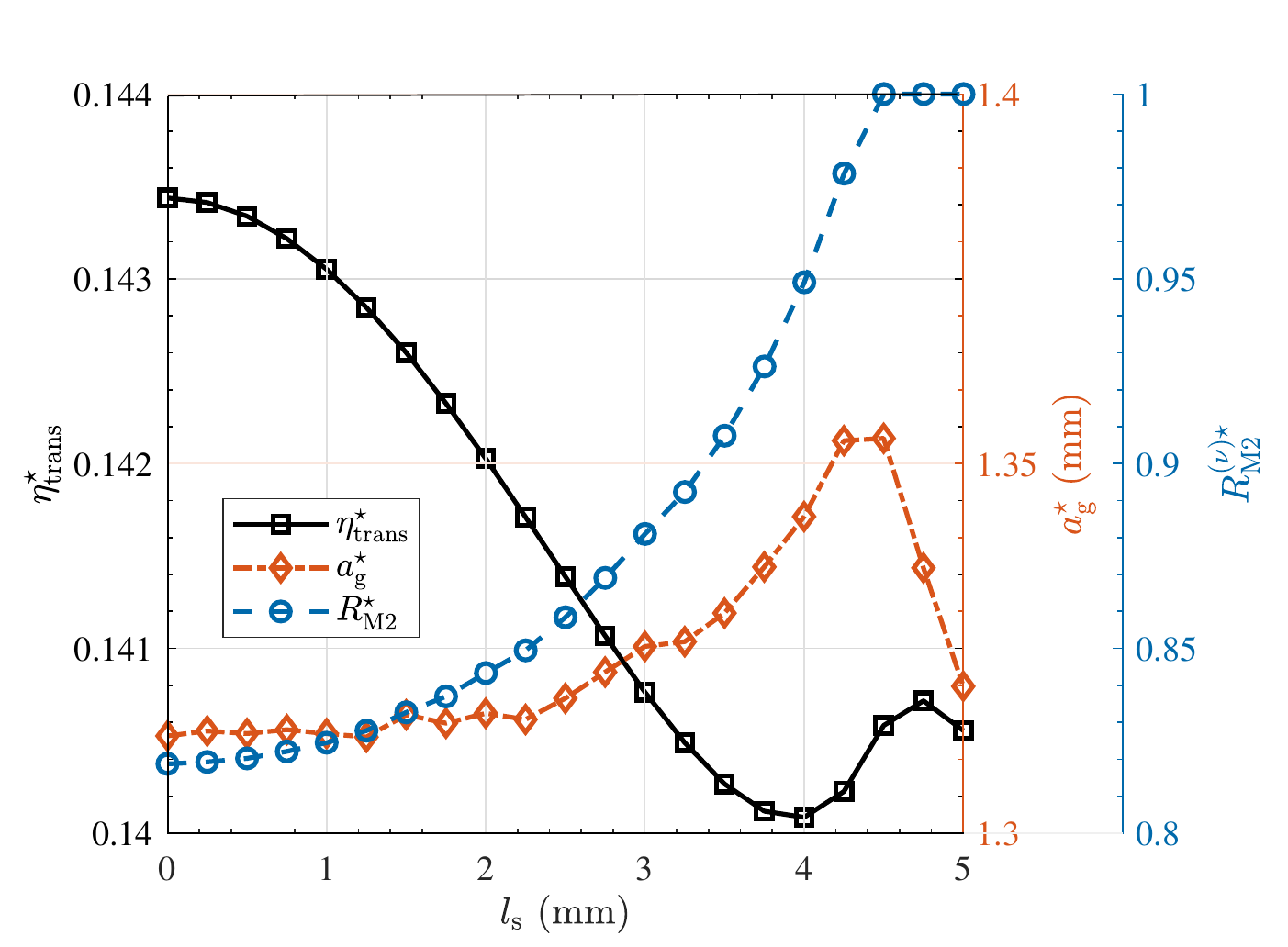}
		\caption{The optimum transmission efficiency $\eta_{\rm trans}^\star$, gain medium radius $a_{\rm g}^\star$, output mirror reflectivity $R_{\rm M2}^{(\nu)\star}$ under different SHG medium thickness $l_{\rm s}$  (The driving power $P_{\rm in}=60$~W; the preset distance $d_{\rm m}=4.32$~m; the focal lengths $\{f_1,f_2\}=5$~cm; the mirror-to-lens intervals $\{l_1, l_2\}$ are obtained by the optimization procedure)}
		\label{fig:eta-ag-R-ls}
	\end{figure}

	The pattern of the transmission efficiency $\eta_{\rm trans}$ related to $a_{\rm g}$ and $R_{\rm M2}^{(\nu)}$ is shown in Fig.~\ref{fig:surfc}. We can find that the surface has a maximum efficiency point~$\eta_{\rm trans}^\star$. Therefore, \textrm{P4} is a convex optimization problem. We adopt the stochastic gradient descent~(SGD) algorithm to obtain a series of solutions of \textrm{P4} under different $l_{\rm s}$, as depicted in Fig.~\ref{fig:eta-ag-R-ls}. We can observe that, as $l_{\rm s}$ grows, the optimum gain medium radius $a_{\rm g}^\star$ and the optimum reflectivity $R_{\rm M2}^{(\nu)\star}$ increase concurrently, while $\eta_{\rm trans}^\star$ decreases slowly. The change of $\eta_{\rm trans}^\star$ is small. Hence, we can deem that $l_{\rm s}$ has little effect on the optimum transmission efficiency. The significant change is $R_{\rm M2}^{(\nu)\star}$ which varies from $82\%$ to $100\%$, and this change correspondingly adjusts  the output PT beam power. We can also observe that $a_{\rm g}^\star$ changes little, especially when $l_{\rm s}<2.5$~mm. As $l_{\rm s}=4.5$~mm, the optimum reflectivity $R_{\rm M2}^{(\nu)\star}$ reaches $100\%$, which indicates that the charging power reduces to $0$. As $l_{\rm s}$ continues to increase, $R_{\rm M2}^{(\nu)\star}$ remains at $100\%$, while $a_{\rm g}^\star$  starts to decrease. Readers should be noticed that the SHG model is valid under the assumption that the input is a plane wave with homogeneous intensity distribution. In practice, the inhomogeneity distribution of the transverse intensity and the beam divergence may reduce the SHG efficiency;  therefore, a larger $l_{\rm s}$ than the ideal case can compensate for this reduction.
	
	
	We demonstrate an off-line optimization procedure, as most parameters, including the gain medium radius~$a_{\rm g}$, the reflectivity of the output mirror~$R_{\rm M2}^{(\nu)}$, and the SHG thickness $l_{\rm s}$, can not be changed after fabrication. This procedure is important as these parameters should be determined before fabrication. An on-line optimization with respect to the lens-to-mirror interval $l_1$ and $l_2$ may further improve the system performance. However, an off-line optimization is sufficient only if the application scenario is specified. For example, in a departure hall the transmitter is mounted on the ceiling, $6$~m above the ground, and the receiver is taken $1~$m above the ground. Assuming the FOV of the transmitter is $60^\circ$\cite{a220219.01} and the receiver moves horizontally in the FOV, we can know that the distance between the receiver and the transmitter lies in the range of $[5~\mbox{m},5.77~\mbox{m}]$. Similarly, in a room the allowable distance lies in the range of $[2~\mbox{m},2.3~\mbox{m}]$. The distance change is small in each specific scenario. Hence, it is sufficient to choose a preset distance for each scenario.
	
	\section{Trade-off Between Charging Power and Communication Rate}
	\label{sec:result}
	
	In our SWIPT system, a portion of the power is used to charge the battery at the receiver, while the other part of the power is used for information transfer.  $l_{\rm s}$ affects the SHG efficiency; hence, it determines how much power is converted into the second-harmonic frequency for information carrying. In this section, we discuss the trade-off on the power allocation for wireless charging and  information transfer.
	
	\begin{figure}
		\centering
		\includegraphics[width=3.4in]{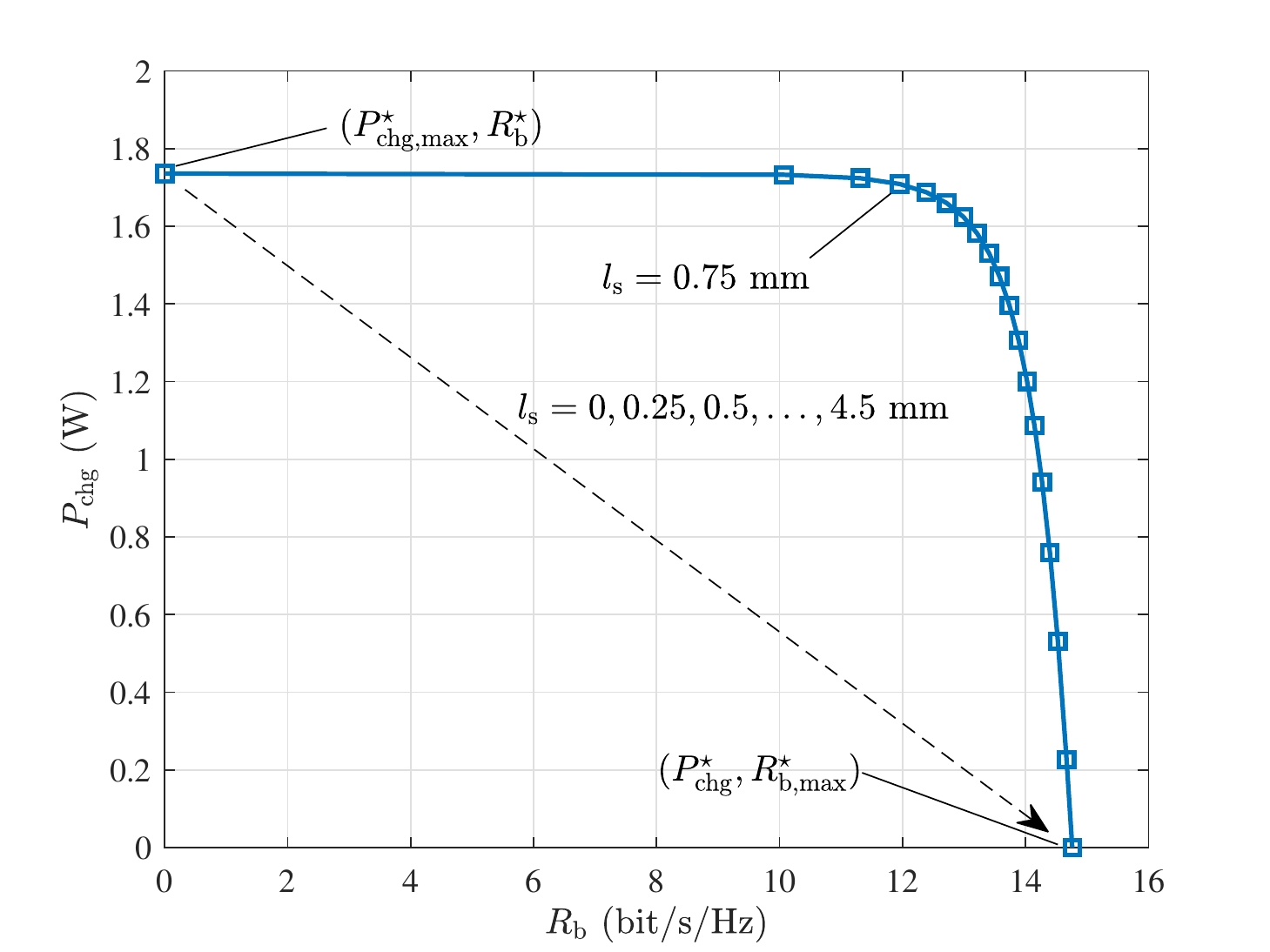}
		\caption{Trade-off on charging power $P_{\rm chg}$ and achievable rate $R_{\rm b}$ (The driving power $P_{\rm in}=60$~W; the preset distance $d_{\rm m}=4.32$~m; the focal lengths $\{f_1,f_2\}=5$~cm; the mirror-to-lens intervals $\{l_1$, $l_2\}$, the reflectivity $R_{\rm M2}^{(\nu)}$, and the gain medium radius $a_{\rm g}$  are obtained by the optimization procedure)}
		\label{fig:tradeoff}
	\end{figure}
	
	All the parameters in this system affect the final charging power and achievable rate, which motivates us to investigate the following question: What is the boundary of the achievable performance? Referring to~\cite{a181126.06}, we define the power-rate~$\mbox{(P-R)}$ region to characterize all the achievable charging power and communication rate pairs ($P_{\rm chg}, R_{\rm b}$), under given input driving power~$P_{\rm in}$ and distance $d_{\rm m}$. 
	By solving \textrm{P4}, we obtain the optimum gain medium radius $a_{\rm g}^\star$ and the optimum output mirror reflectivity $R_{\rm M2}^{(\nu)\star}$. Now, we choose different SHG medium thickness ($l_{\rm s}=0,  0.25, 0.5, \dots, 4.5$~mm) and obtain the optimum parameters $(a_{\rm g}^\star, R_{\rm M2}^{(\nu)\star})$ for each case. Then, we use these optimum parameters to calculate the optimum charging power $P_{\rm chg}^\star$ and the optimum rate $R_{\rm b}^\star$ under  distance $d=d_{\rm m}$. The relation between $P_{\rm chg}^\star$  and $R_{\rm b}^\star$ is demonstrated in Fig.~\ref{fig:tradeoff}. This curve shows the boundary of the achievable ($P_{\rm chg}, R_{\rm b}$) pairs under this distance. The bottom-left side of the boundary is the achievable P-R region. It is easy to identify two boundary points denoted by ($P_{\rm chg,max}^\star, R_{\rm b}^\star$) and  ($P_{\rm chg}^\star, R_{\rm b,max}^\star$), which show the maximum achievable charging power and communication rate, respectively.
	However, extremely high rate in bit/s/Hz is not useful, as the modulator accuracy is hard to support such high modulating order. On the other hand, the charging power should be as high as possible. Therefore, for practical use, the SHG medium thickness $l_{\rm s}$ can be set as $0.75$~mm to obtain a very high charging power close to the maximum achievable value $P_{\rm chg,max}^\star$ while keeping the rate high enough in use.

\begin{figure}
		\centering
		\includegraphics[width=3.4in]{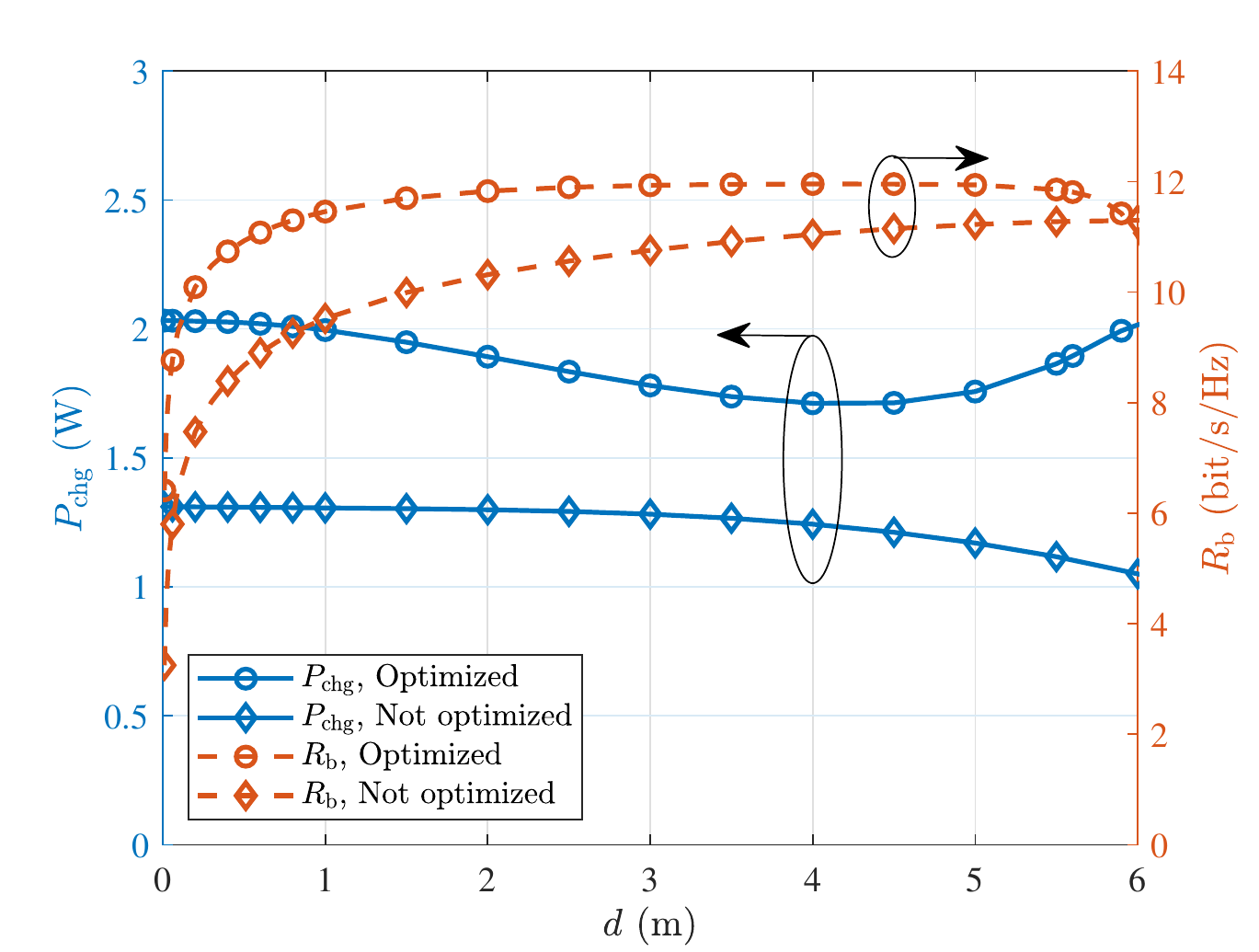}
		\caption{Performance comparison between the optimized asymmetric system in this work and the non-optimized symmetric system  in the previous work (The driving power $P_{\rm in}=60$~W, other parameters in this work are obtained by the presented optimization procedure)}
		\label{fig:perform-cmp}
	\end{figure}

	Next, we compare the performance of the optimum design with the previous work presented in~\cite{MXiong2021.SWIPT}. We set $P_{\rm in}=60$~W.  Then, we solve \textrm{P4} to obtain the optimum functional parameter tuple; that is $\mathbf{m}^\star=(1.31~\mbox{mm},  82.2\%)$. According to the boundary of the P-C region, we decide that $l_{\rm s}=0.75$~mm.  Using these optimum parameters, we calculate the  charging power $P_{\rm chg}$ and the achievable rate $R_{\rm b}$ under different transmission distance $d$, as depicted in~Fig.~\ref{fig:perform-cmp}. As the distance is close to zero, we can see that the capacities in the two cases drop quickly. This is because the beam radius at the SHG medium becomes very large when the distance is close to zero so that the SHG efficiency is very small. We can observe that the optimized asymmetric system exhibits  better performance than the non-optimized symmetric system  in~\cite{MXiong2021.SWIPT}. For $6\mbox{-m}$ distance, the charging power of the optimized system is improved by $91.8\%$, compared with the non-optimized symmetric system. The rate is kept above $11.3$~bit/s/Hz for a large range of distance, which is also superior to\mrr{} the performance of the non-optimized system. Since a UAV generally needs tens-of-watts charging power, the power provided by the demonstrated system can only slightly expand the duration of flight. The achievable power is sufficient for portable electronic devices, such as smart watches, smart phones, electric toothbrushes. In the future, with the development of new pump modules, gain medium modules, and photovoltaic materials, the charging power has the potential to be improved, according to the analysis model. Besides, using multiple transmitters can also improve the total charging power.

	\section{Conclusions}
	\label{sec:con}
	
	In this paper, we investigated a mobile optical simultaneous wireless information and power transfer~(SWIPT) system based on  asymmetric spatially separated laser resonator~(SSLR) and intra-cavity second harmonic generation~(SHG). We created the analysis model and presented the optimization procedure for parameter determination.
	From the results, we found that the focal lengths of  lenses in focal cat's eye retroreflectors~(FTCRs) are not the crucial factors that affect the performance, but the lens-to-mirror intervals of FTCRs and some other parameters, including the gain medium radius, the reflectivity of the output mirror, and the SHG medium thickness, need to be determined by optimization algorithm.
	Numerical results show that the SWIPT performance in this work is greatly improved, compared with the symmetric-SSLR-based SWIPT system proposed in the previous work. Besides, we investigated the power-rate~(P-R) region to demonstrate all the achievable pairs of charging power and communication rate. The boundary of the P-R region depicts the optimum performance and gives a guideline on the trade-off between power transfer and information transfer. With the trade-off strategy, the system model becomes more flexible, as it can be easily switched  into a pure communication system, a pure wireless charging system, or a SWIPT system.


	
	%

	%



	\ifCLASSOPTIONcaptionsoff
	\newpage
	\fi

	
	
	
	\bibliographystyle{IEEETran}
	\small
	%
	\bibliography{mybib}
	%
	%
	
	%
	
	%
	%
	
	
	
	
	

	
\end{document}